\documentclass{elsart}

\usepackage{graphics,amssymb}

\begin{document}

\begin{frontmatter}




\title{Creation of twistless circles in a model of stellar pulsations}

\author[label1]{Andreea  Munteanu},  
\author[label2]{Emilia Petrisor},
\author[label1,label3]{Enrique Garc\'\i a-Berro} \&
\author[label3,label4]{Jordi Jos\'e}

\address[label1]{Departament  de   F\'{\i}sica  Aplicada,  Universitat
		 Polit\`ecnica de Catalunya, Jordi Girona Salgado s/n,
		 M\`odul  B--5, Campus  Nord, 08034  Barcelona, Spain}
\address[label2]{Departamentul     de     Matematica,    Universitatea
		 Politehnica  Timisoara,  Pta  Regina  Maria  1,  1900
		 Timisoara, Romania}
\address[label3]{Institut  d'Estudis Espacials  de  Catalunya, Edifici
		 Nexus, Gran Capit\`a 2-4, 08034 Barcelona, Spain}
\address[label4]{Departament  de  F\'{\i}sica  i  Enginyeria  Nuclear,
		 Universitat  Polit\`ecnica de  Catalunya,  Av.  V\'\i
		 ctor  Balaguer, s/n,  08800, Vilanova  i  la Geltr\'u
		 (Barcelona), Spain}

\begin{abstract}
In the present paper, we study the  Poincar\'{e}  map  associated to a
periodic perturbation, both in space and time, of a linear Hamiltonian
system.  The  dynamical  system  embodies  the  essential  physics  of
stellar  pulsations and provides a global and qualitative  explanation
of the chaotic oscillations observed in some stars.  We show that this
map is an area  preserving  one with an  oscillating  rotation  number
function.  The  nonmonotonic  property of the rotation number function
induced by the  triplication of the elliptic fixed point is superposed
on the  nonmonotonic  character due to the  oscillating  perturbation.
This superposition leads to the  co-manifestation of generic phenomena
such as reconnection and meandering,  with the nongeneric  scenario of
creation  of   vortices.  The   nonmonotonic   property   due  to  the
triplication  bifurcation is shown to be different from that exhibited
by the cubic  H\'{e}non  map, which can be considered as the prototype
of area preserving  maps which undergo a triplication  followed by the
twistless bifurcation.  Our study exploits the reversibility  property
of the initial system, which induces the time-reversal symmetry of the
Poincar\'{e} map.
\end{abstract}


\end{frontmatter}

\section{Introduction}

During the past twenty years there has been increasing interest in the
study of the  coexisting  regular and chaotic  motions in  Hamiltonian
systems with few degrees of freedom.  A basic problem of study was the
transition  to chaos in two  degrees of  freedom  Hamiltonian  systems
which  are  slight   perturbations  of  an  integrable  system,  whose
Hamiltonian is nondegenerate.  In order to ease this task, most of the
results on the  dynamics of such  Hamiltonian  systems  were  obtained
exploring the behaviour of a  Poincar\'{e}  map  associated to a cross
section  in a fixed  energy  surface.  A  Poincar\'{e}  map is an area
preserving  map (APM)  defined on an annulus,  that is,  depending  on
$(r,\theta)$, where $r$ is a momentum--like coordinate and $\theta$ is
an  angular  variable.  Considering  the  lift of  such a map,  namely
computing  the  angle  without  the  restriction  modulo  $2\pi$,  one
associates to an orbit $(r_n,  \theta_n)$ the rotation number which is
the limit:

\begin{equation}
\rho = \lim_{n\rightarrow\infty}\frac{\theta_n-\theta_{0}}{2\pi n},
\label{eq:rotation} 
\end{equation}

\noindent if it  exists.  If $\rho$ is an  irrational number, then the
orbit densely fills  an invariant circle (KAM circle),  while an orbit
of rational  rotation number $p/q$  --- with $p, q$  coprime integers,
$q>0$ --- is a $q$--periodic orbit.

Typical questions  of mathematical  and physical interest  include the
persistence of the KAM circles after perturbation and determination of
the threshold at which a circle  of a given rotation number breaks up.
The  main  results which  answer  these  questions,  such as  the  KAM
theorem,  the  Poincar\'{e}--Birkhoff  theorem,  and the  Moser  twist
theorem, are based  on the validity of the twist  property of the area
preserving map transforming $(r,\theta)\mapsto (r',\theta')$, namely:

\begin{equation}
\frac{\partial \theta'}{\partial r} \neq 0,\,\,\forall\,\, r
\end{equation}

\noindent  The twist  condition implies  the monotonic  change  of the
rotation number $\rho$ with $r$.  The APM satisfying this condition is
called monotonic  twist map  or simply a  twist map.  During  the last
twenty  years  the  monotonic  twist  APMs  were  studied  extensively
\cite{M92}.  Consequently,  their dynamics has  gained almost complete
understanding.  In particular, this  statement is a consequence of the
late understanding  that Hamiltonian systems  have universal behaviour
which  can be  revealed by  studying explicit  symplectic maps  of the
plane.

Recently, due to the increasing number of physical  phenomena that are
modeled by nontwist maps, this class of dynamical  systems has finally
captured the attention of the  scientific  community.  The few studies
on  nontwist  maps  existent  in  the  literature   have  revealed  an
unexpected variety of local and global bifurcations:  dimerized island
chains, periodic orbits  collision,  separatrix  reconnection,  vortex
islands \cite{C-NEA97,SA97,P01}.

In the present paper we study the  Poincar\'{e}  map  associated  to a
periodic  perturbation,  both  in  space  and  in  time,  of a  linear
Hamiltonian  system.  The dynamical system  originates from a model of
stellar pulsations previously  introduced in \cite{IEA92} for low mass
stars  ($M/M_\odot  < 8$,  where  $M_\odot$  is the  solar  mass)  and
extended  to  intermediate  mass stars  ($M/M_\odot  \in  [8,11]$)  in
\cite{MEA02}.  The  latter  work  included  some  preliminary  results
concerning the nontwist character of the map.  Here, we show that this
map is an APM with an oscillating  rotation number function.  In spite
of the system being a perturbation of a linear  Hamiltonian,  we argue
that due to both the  acquired  nontwist  character  and the  periodic
character of the  perturbation,  the map presents the typical features
of a  generic  class  of  nontwist  APMs,  such  as  reconnection  and
meandering, with the nongeneric scenario of creation of vortices.  The
nonmonotonic property due to the triplication  bifurcation is shown to
be different from that exhibited by the cubic H\'{e}non map, which can
be  considered  as the  prototype  of  APMs,  and  which  undergoes  a
triplication   followed  by  the  twistless   bifurcation.  Our  study
exploits  the  reversibility  property of the  initial  system,  which
induces the time-reversal symmetry of the Poincar\'{e} map.  The paper
is organized  as follows.  In \S 2 we  introduce  our model of stellar
pulsations.  Before  getting  into  the  detailed  analysis  of it, we
consider worth recalling the most important features of nontwist maps.
This is done in \S 3.  In this  section  we will  also  emphasize  the
consequences of both the triplication  bifurcation and of the symmetry
properties.  Later, in \S 4 we study the dynamics of our simple  model
of stellar  pulsations.  Finally in section 5 we  summarize  our major
findings and we draw our conclusions.

\section{Description of the model}

Pulsating  stars are perfect tests of the theory of stellar  evolution
since   the   comparison    between   the   theoretical    pulsational
characteristics  (periods, amplitudes, and growth time scales) and the
observed ones can help substantially in the fine tuning of the stellar
models.  Low- and  intermediate-mass  stars  are  prone to  experience
recurrent  thermal  instabilities  and  substantial  mass loss  during
certain intervals of their evolution.  Moreover, a substantial  amount
of pulsating  stars in the Galaxy shows  irregular  behaviour and this
makes them interesting  targets in the field of dynamics systems.  The
present  model  embodies  the  essential  of stellar  oscillations  by
assuming an oscillatory driving originating in the stellar interior in
the form of  sinusoidal  pressure  waves.  This model  belongs  to the
family of the  so-called  one-zone  models  which  treat the star as a
rigid  core  surrounded  by a  dynamic,  homogeneous  gas  shell.  The
pressure waves propagate through a transition zone  characterized by a
certain  transmission  coefficient until they hit the stellar envelope
and dissipate.  No back reaction of the outer layers on the inner ones
is  considered.  The  dissipation  of the  pressure  waves  induces  a
fluctuation in the radius and velocity of the outermost layer and this
variability makes the object of our study.

The  variation  of  the   interior  radius,  $R_{\rm  c}$,  around  an
equilibrium value, $R_0$, is  given by $R_{\rm c}=R_0+\alpha R_0 ~\sin
~\omega_{\rm c} \tau$, where $\tau$  is the time in years and $\alpha$
and $\omega_{\rm c}$ are the fractional amplitude and the frequency of
the  driving, respectively.   In  absence of  any  driving force,  the
equation of motion is

\begin{equation}
\frac{d^2 R}{d\tau^2} = \frac{4\pi R^2}{m}P -\frac{GM}{R^2},
\label{eq:motion} 
\end{equation}

\noindent where $M$ is the stellar mass and $R$ and $P$ are the radius
and pressure in the shell of mass $m$, respectively.  For convenience,
we introduce the nondimensional variables $x \equiv R /R_\star -1$ and
$t \equiv \omega_{\rm m}\tau$, where $R_\star$ is the equilibrium stellar
radius and

\[
\omega_{\rm m} \equiv \sqrt{\frac{GM}{R_\star^3}}
\] 
\noindent is the characteristic frequency of the star.

Considering   that  the   additional   perturbative  acceleration   is
proportional  to  the  driving  acceleration  with  a  proportionality
coefficient $Q$ we obtain

\begin{equation}
A ~=~ -Q \left.\frac {d^2 R_c}{d\tau^2} ~\right|_{\rm ret}
\label{eq:acceleration}
\end{equation}

\noindent and after some algebraic manipulation \cite{MEA02} the final
equation of motion reads

\begin{equation}
\begin{array}{lll}
\dot x&=&y \\ 
\dot{y} &=& -x-\epsilon~ \sin (\omega x-\omega t -\alpha\omega ^{1/3} 
\sin ~\omega t )~ ,
\end{array}
\label{eq:motion2}
\end{equation}

\noindent where  the characteristic frequency of  the system, $\omega$
results  from the  use of  the dimensionless  time unit  and  from the
assumption that $R_0$ encompasses  almost the entire stellar mass.  It
is defined as

\begin{equation}
\omega  \equiv \frac{\omega_{\rm  c}}{\omega_{\rm m}}  =  \left( \frac
{R_0}{R_ \star}\right) ^{-3/2}. \label{eq:omega}
\end{equation}

In the action-angle coordinates $(r,\theta)$ resulting from the change
$x = \sqrt  {2r} \cos \theta$ and  $ y = \sqrt {2r}  \sin \theta$, the
Hamiltonian of the system is

\begin{equation}
H(r,\theta) = r - \frac{\epsilon}{\omega}~ \cos 
(\omega\sqrt{2r}\cos\theta-\omega t-\alpha\omega^{1/3}\sin\omega t)
\label{eq:Ham}.
\end{equation}

The  parameters  of the  system  are  $\epsilon$,  the  total  driving
amplitude,  $\alpha$, the fractional  driving  amplitude and $\omega$,
the characteristic frequency of the system.  From the point of view of
the  parametric  study,  we take  $\omega$  to be the  most  important
parameter  fixing the type of star to be studied and its  evolutionary
status.  It is a measure  of the ratio  between  the  internal  radius
where the driving is  produced  and the  equilibrium  stellar  radius.
From the astrophysical  perspective, we are interested in the dynamics
of the system  from  Eq.(\ref{eq:motion2})  corresponding  to  $\omega
\approx 3$, more exactly  $\omega=3.0146$.  This low value of $\omega$
is associated  with evolved stages of  intermediate  mass stars, as it
results from stellar evolutionary models \cite{REA96,GEA97}.  Since we
are interested in modeling the evolution of this type of stars, we fix
$\omega$  during  our   simulations.  We  have  performed  a  thorough
parametric  study in the space  ($\epsilon$,$\alpha$)  as it assures a
fine tuning of the strength of the  perturbation.  As in \cite{MEA02},
we are  interested  in $\epsilon < 100\% $ and $\alpha < 40 \%$, which
allow the perturbative approach.

\section{Nontwist maps}

The  prototype  of nontwist  area  preserving  maps  is the  quadratic
standard map \cite{P01,C-NGM96}:

\begin{equation}
\begin{array}{lll}
r_{n+1}&=&r_{n}-k\sin{\theta}\\
\theta_{n+1}&=&\theta_n+2\pi\omega-r_{n+1}^2\,\,(\mbox{mod}\,\,
2\pi)~.
\end{array}
\label{eq:quadratic}
\end{equation}

The  twist condition  is violated  along the  curve $r=k\sin{\theta}$.
For $k=0$ we get an integrable map, whose orbits lie on the circles of
constant   $r$.    The   rotation   number   function   is   $\rho(r)=
\omega-r^2/(2\pi)$. The  circle $r=0$ has the  maximum rotation number
$\rho_{\max}=\omega$. It  is called the twistless  circle or shearless
circle.   A  slight  perturbation  leads  to the  persistence  of  the
twistless  circle and  nearby  circles having  a diophantine  rotation
number  ---  that  is,  an  irrational number  badly  approximated  by
rationals \cite{S98}. Hence, the nontwist standard map, defined above,
has an invariant circle of  maximum rotation number among the rotation
numbers of the nearby orbits.   For a fixed perturbation parameter $k$
and for  $\omega$ chosen  in such  a way that  the map  has a  pair of
Poincar\'{e}--Birkhoff chains  containing $p/q$-type  periodic orbits,
as  $\omega$  varies the  two  chains approach  each  other  and at  a
threshold $\omega=\omega_r(k)$ the hyperbolic points of the two chains
enter the twistless circle.  This is called the reconnection threshold
\cite{HH95}.    At  the  reconnection   threshold  the   two  distinct
hyperbolic  orbits,  of  rotation   number  $p/q$,  are  connected  by
heteroclinic arcs.

\begin{figure}[t]
\centerline{\scalebox{0.55}{\includegraphics{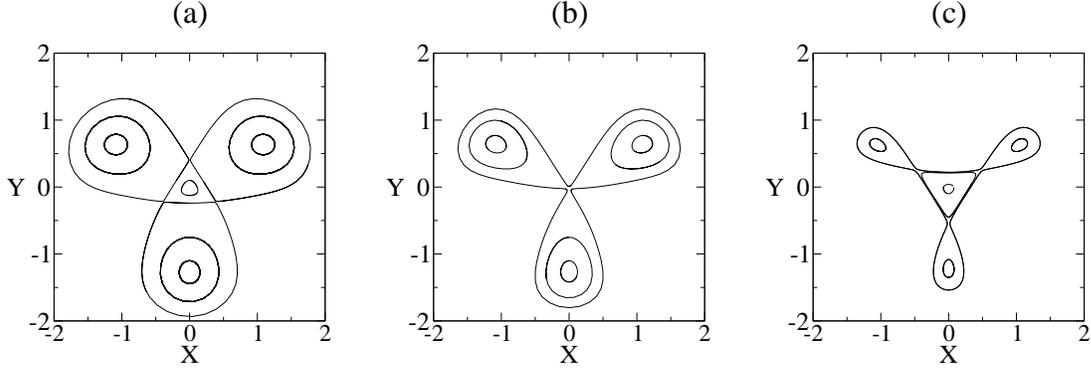}}}
\caption{The  triplication  scenario.  The phase  portrait  around the
	 elliptic point at the triplication threshold is shown in {\sl
	 (b)}.  Observe the change in the position of the heteroclinic
	 triangle in {\sl (a)} and {\sl (c)}.}
\label{tripl}
\vspace{0.5cm}
\end{figure}

There are many ways in which  the  monotonic  twist  condition  can be
violated.  The  vertical  lines  $\theta=\theta_0$  can be mapped into
curves having a single extremum --- quadratic twist \cite{C-NGM96} ---
or multiple extrema --- cubic  \cite{C-NF02},  quartic  \cite{HH95} or
sinusoidal   twist   \cite{SEA97}.  Moreover,   it  has  been   proved
\cite{DMS00}  that whenever the elliptic  fixed point of an APM of the
plane passes through a triplication,  a twistless  bifurcation  occurs
or,  equivalently,  the rotation  number as a function of the distance
from the elliptic  fixed point becomes a  nonmonotonic  function.  The
triplication  of the  elliptic  fixed  point of an APM  $f_\mu$ of the
plane, occurs at the value $\mu \equiv \mu_3$ at which the multipliers
of the elliptic point cross the values $\lambda=\exp(\pm{2\pi  i/3})$.
At the triplication  threshold $\mu=\mu_3$ an unstable period--3 orbit
emerges from the elliptic fixed point in both directions,  that is for
$\mu<\mu_3$  and for  $\mu>\mu_3$.  As $\mu$  increases  the  elliptic
point  crosses the $1/3$  resonance  and the  triangular  shape figure
changes  side  after   shrinking  to  elliptic   fixed  point  (Figure
\ref{tripl}).

The symmetry  properties are  very useful for  the explanation  of the
behaviour exhibited  by the system  under study.  We next  recall some
notions and results concerning  time--reversal symmetry of a dynamical
system \cite{L98}. A system of differential equations

\begin{equation}
\dot{x}=F(x), x\in \mathbb{R}^n,
\end{equation}

\begin{figure}[t]
\hspace{-0.5cm}
\centerline{\scalebox{0.43}{\includegraphics{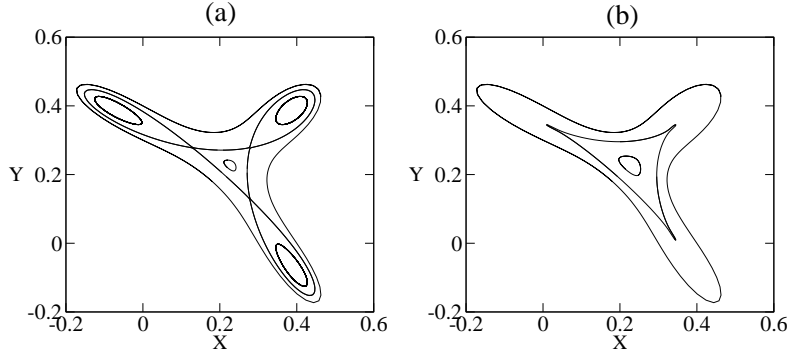}}}
\caption{The  triplication  bifurcation  in the cubic  H\'{e}non  map.
	 {\sl  (a)}  The  separatrix   associated  to  the  hyperbolic
	 period--3   orbit  of  the  map   from   Eq.(\ref{eq:henon}),
	 corresponding  to  $\mu=-1.13$,  and some nearby orbits; {\sl
	 (b)} The same map at the saddle--center  threshold  $\mu_{\rm
	 s}=-1.12189$.  The twistless  circle is the circle with three
	 cusps.}
\label{hentripl}
\vspace{0.5cm}
\end{figure}

\noindent or  equivalently, the vector field $F$  is called reversible
if there  is a smooth  involution $R:\mathbb{R}\to\mathbb{R}$ ($R\circ
R={\rm{Id}}$)  such that  $DR\circ  F=-F\circ R$,  where  $DR$ is  the
differential of the map $R$. $R$  is called the reversor of the vector
field $F$.  If $\Phi_t$ is the  flow of the vector field $F$, then the
reversibility means  that $\Phi_{-t}R=R\Phi_t$. This  is tantamount to
say that the reflection of  the trajectory $x(t)=\Phi_tx_0$ is also an
orbit of the  system. A diffeomorphism $f:\mathbb{R}^2\to\mathbb{R}^2$
is  called reversible  with  respect  to a  smooth  involution $R$  of
$\mathbb{R}^2$   if  $f^{-1}=R\circ   f\circ   R$.   If   $f$  is   an
$R$-reversible diffeomorphism then $I=f\circ  R$ is also an involution
$(I\circ I=\rm{Id})$, and $f=I\circ  R$. This factorization of the map
$f$ is very useful for the  study of its dynamical properties.  It can
be  shown  that  $f^n$  is  a $R$-reversible  map,  too,  $\forall\,\,
n\in\mathbb{Z}$.  Denote  by $I_n=f^n\circ R$,  $n\in \mathbb{Z}$, and
by $\Gamma_n$ the fixed point  set of the involution $I_n$. $\Gamma_n$
are called symmetry lines of the map $f$.  A point in the intersection
$\Gamma_j\cap\Gamma_k$  is a  periodic  point of  the  map $f$,  whose
period divides  $\vert j-k\vert$.   The symmetry lines  $\Gamma_k$ are
transformed by $f^n$ into other symmetry lines in the following way:

\begin{equation}
\label{symmlin}
f^n\Gamma_k=\Gamma_{2n+k}
\end{equation}

In  order  to interpret  the  nontwist--type  dynamics  which will  be
encountered  in  our system  when  studying  the  triplication of  the
elliptic  fixed  point,  we   consider  necessary  to  illustrate  the
twistless bifurcation in the case of the cubic H\'{e}non map

\begin{equation}
\begin{array}{lll}
x_{n+1}&=&-y_n+x_n^3+\mu x_n+0.7 \\ y_{n+1}&=& x_n~,
\end{array}
\label{eq:henon}
\end{equation}

\noindent which  has an elliptic fixed point  $(x_0,y_0=x_0)$ that can
be  easily   found.   At   $\mu_3=-1.1505$  this  point   undergoes  a
triplication,  while  at  $\mu_{\rm  s}=-1.12189$  the  saddle  center
collision  occurs. In  Figure \ref{hentripl},  we show  that  as $\mu$
increases beyond $\mu_3$, the  elliptic and hyperbolic period-3 orbits
collide and disappear in a saddle--center bifurcation at $\mu = 
\mu_{\rm s}$.

\begin{figure}[t]
\centerline{\scalebox{1.0}{\includegraphics{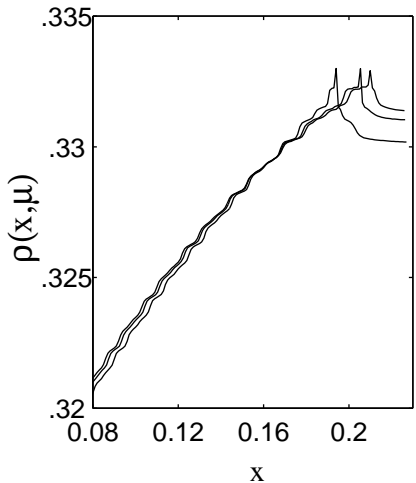}}}
\vspace{-0.2cm}
\caption{Rotation  number $\rho(x,\mu)$ versus $x$ for three values of
	 $\mu$ between the triplication and  saddle--center  collision
	 thresholds of the  H\'{e}non  cubic map  (Eq.\ref{eq:henon}).
	 The  curves  correspond  to  $\mu=-1.135$,   $\mu=-1.13$  and
	 $\mu=\mu_{\rm s}=-1.12189$.  See text for more details.}
\label{henrot1}
\end{figure}

In order to better identify the twistless bifurcation --- the birth of
a twistless  circle --- in Figure  \ref{henrot1}  we show the rotation
number  $\rho(x,\mu)$  as a function  of the  points  $x$ lying on the
symmetry line  $\Gamma_0:=\{(x,y) | y=x\}$ of the cubic H\'{e}non map,
which is a reversible map with respect to the reversor $R(x,y)=(y,x)$.
More  precisely, the rotation  number is computed at the points $x \in
\Gamma_0$ and  satisfying  $x<x_0$.  Figure  \ref{henrot1}  shows that
between  the  triplication  threshold  and  saddle--center   collision
threshold of the two  period--3  orbits, the  twistless  circle is the
circle of rotation  number  $\rho_{\rm  max}=1/3$  (the maximum points
from  right  to left  correspond  to  $\mu=-1.135$,  $\mu=-1.13$,  and
$\mu=\mu_{\rm  s}=-1.12189$).  Observe that as $\mu$  increases  up to
$\mu_{\rm s}$, the twistless circle moves away from the elliptic fixed
point.

\begin{figure}[t]
\centerline{\scalebox{1.0}{\includegraphics{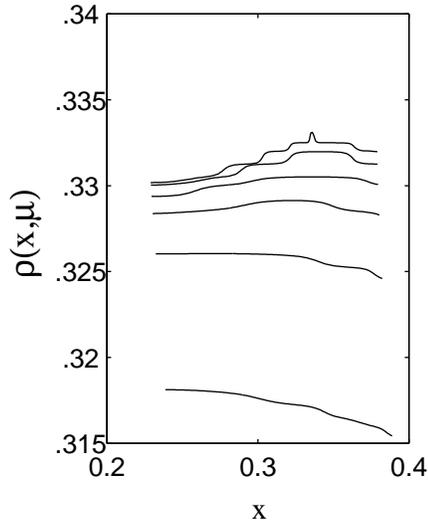}}}
\vspace{-0.2cm}
\caption{Rotation  number  $\rho(x,\mu)$  of the points $(x,y)$ on the
	 symmetry line $\Gamma_0$ of the map $f_\mu$, where $f_\mu$ is
	 the  H\'{e}non  map.  From  top  to  bottom:  $\mu=-1.12189$,
	 $\mu=-1.12$,   $\mu=-1.115$,    $\mu=-1.105$,    $\mu=-1.08$,
	 $\mu=-1$ and for $x>x_0$,  where $(x_0, y_0)$ is the elliptic
	 fixed point.}
\label{henrot2}
\end{figure}

Much more important  (and  difficult) is to detect the location of the
twistless  circle  after  the   collision--annihilation   of  the  two
period--3 orbits.  In order to get some insight into the manifestation
of the twist  propertiy  beyond the threshold  $\mu_{\rm  s}$, we have
computed numerically the rotation numbers for points starting near the
elliptic  fixed point of the cubic  H\'{e}non  map and on the symmetry
line  $\Gamma_0$.  The rotation  number  functions  $\rho(x,\mu)$  are
plotted for  $\mu=\mu_{\rm  s}$, and some  parameter  values  obtained
varying  $\mu$  in the  direction  given  by  the  vector  $v=\mu_{\rm
s}-\mu_3$  (Figure  \ref{henrot2}).  Note that the cubic H\'{e}non map
has a twistless  circle of maximum  rotation number among the rotation
numbers of nearby  orbits.  Varying the  parameter  $\mu$ as specified
above,    $\rho_{\rm    max}$   decreases   as   a   result   of   the
collision--annihilation  of the orbits  \cite{P01}, and at some value,
$\mu_{\rm t}$, the rotation number becomes  decreasing.  At this point
the map is no longer  nontwist.  This is the reason why  $\mu=\mu_{\rm
t}$  was  called  the  threshold  of  the  twistless   bifurcation  in
\cite{DMS00}  where the above presented  scenario was first  presented
although in the reverse order.

\section{Dynamical properties of the Poincar\'e map}
 
Due to the the complexity  of the perturbative term of the Hamiltonian
of Eq.(\ref{eq:Ham}),  the associated system  in Eq.(\ref{eq:motion2})
is analytically  intractable.  In order  to get some insight  into its
dynamical  behaviour,  we study  the  Poincar\'{e}  map  (in fact  the
stroboscopic map) associated to this time-periodic Hamiltonian system.
The unperturbed Hamiltonian $ H_0(r,\theta)=r$ is globally degenerate,
which means that ${\partial^2  H_0}/{\partial r^2}=0, \,\, \forall r$.
The dynamical  consequence of this  degeneracy is that all  the orbits
are  periodic having  the same  period.  This  is in  contrast  to KAM
theory, where  the unperturbed system is nondegenerate.   We will show
that after a periodic perturbation (periodic in time and in space) the
corresponding Poincar\'{e} map  exhibits local and global bifurcations
which in  some cases are typical  of the class of  nontwist APMs which
are perturbations of locally degenerate integrable APMs of an annulus,
while in other cases are not.

\subsection{The nontwist property of the map}

Since the  system given  in Eq.(\ref{eq:motion2}) is  time-periodic of
period  $T=2 \pi/\omega$,  we  define the  function $s:\mathbb{R}  \to
\mathbb{S}^1$,    $s(t)=t\,\,(\mbox{mod}\,\,    2\pi/\omega)$,   where
$\mathbb{S}^1$   is   the   circle   identified  with   the   interval
$[0,2\pi/\omega)$.    Hence,  we   have  the   autonomous   system  of
differential equations:

\begin{equation}
\begin{array}{lll}
\dot x&=&y \\     
\dot{y}&=&-x-\epsilon~\sin(\omega x-\omega s-\alpha\omega^{1/3} 
\sin ~\omega s )\\ 
\dot{s}&=&1~.
\end{array}
\label{eq:motionf}
\end{equation}

\noindent We denote by $\Phi^\epsilon_t$  its flow.  $\Phi^\epsilon_t$
associates   to  each   triplet   $(x_0,y_0,s_0)\in   \mathbb{R}\times
\mathbb{S}^1$  the  position  at the  time  moment  $t$ of  the  orbit
starting at $t=0$ from $(x_0,y_0,s_0)$.  The plane

\begin{equation}
\Sigma  ~=~ \{(x,y,s)  \in  \mathbb{R}^2 \times  \mathbb{S}^1 |  s=0\}
\equiv \mathbb{R}^2
\end{equation} 

\noindent is transversal to the  flow and the map $P_\epsilon : \Sigma
\rightarrow     \Sigma$    defined    by     $P_\epsilon    (x,y)=\Phi
_{2\pi/\omega}^\epsilon (x,y,0)$ is the associated Poincar\'e map.

It  is a  classical  result  in  dynamical  systems  theory  that  the
Poincar\'{e}  map  associated  to a one  degree  of  freedom  periodic
time--dependent  Hamiltonian  system is an APM.  In order to interpret
the dynamics displayed by the numerically  computed  Poincar\'{e} map,
let us deduce  some  geometrical  properties  of this map.  The vector
field $F$, associated to the system from Eq.(\ref{eq:motionf}),

\begin{equation} 
F(x,y,s)=(y,-x-\epsilon\sin(\omega x-\omega s- b\sin(\omega s)), 1),
\end{equation} 

\noindent with ($b=\alpha\omega^{1/3}$), is reversible with respect to
the involution  $\mathcal{R}:  \mathbb{R}^2  \times  \mathbb{S}^1  \to
\mathbb{R}^2    \times    \mathbb{S}^1$,     $\mathcal{R}(x,y,s)     =
(-x,y,2\pi/\omega-s)$, that is:

\begin{equation}
\label{revvf}
F\circ \mathcal{R}=-D\mathcal{R}\circ F.
\end{equation}

The   reversibility   property    (\ref{revvf})   is   equivalent   to
$\mathcal{R}\circ     \Phi^\epsilon_{-t}    =     \Phi^\epsilon_t\circ
\mathcal{R}$,  $\forall  t\in\mathbb{R}$,  which  means that  $\forall
t\in\mathbb{R}$     the    diffeomorphism     $\Phi^\epsilon_t$     is
$\mathcal{R}$--reversible,         because        $\Phi^\epsilon_{-t}=
(\Phi^\epsilon_t)^{-1}$.  The  Poincar\'{e}   map  associated  to  the
vector field $F$ is  $R$--reversible,  with respect to the  involution
$R:\Sigma\to\Sigma$,   $R(x,y)=(-x,y)$.  Actually,   notice  that  the
reversor of the vector field $F$ is related to the involution  $R$ by:

\begin{equation}
\mathcal{R}(x,y,s)=(R(x,y),2\pi/\omega-s).
\end{equation}

It is  clear that  the definition of  reversible maps is  fulfilled by
$P_\epsilon$:

\begin{equation}
\begin{array}{l}
P_\epsilon(R(x,y))=\Phi_{2\pi/\omega}(-x,y,0)=\Phi_{2\pi/\omega}
(\mathcal{R}(x,y,2\pi/\omega))=                                                  \\
\mathcal{R}\Phi_{-2\pi/\omega}(x,y,2\pi/\omega)=RP_\epsilon^{-1}(x,y).
\end{array}
\end{equation}

The symmetry line useful in the analysis of the dynamical behaviour of
the Poincar\'{e} map  is $\Gamma_0=\mbox{Fix}(R)$, having the equation
$x=0$.

\begin{figure}[t]
\centerline{\scalebox{0.55}{\includegraphics{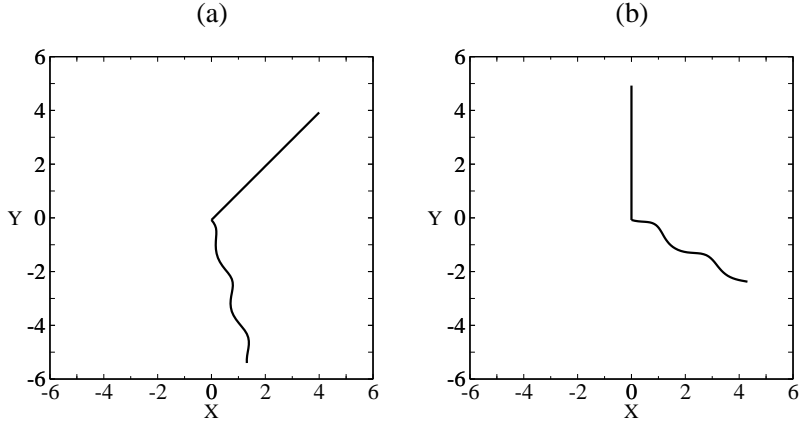}}}
\caption{The nontwist property of the Poincar\'{e} map.  {\sl (a)} The
	 effect of the map on the semi-line defined by $\theta=\pi/4$;
	 {\sl  (b)}  The  effect  of  the  map  on  the  semi-line  of
	 $\theta=\pi/2$.}
\label{semilines}
\vspace{0.5cm}
\end{figure}

The  Poincar\'{e}  map $P_0$ has the  elliptic  fixed  point  $(0,0)$.
After a slight  perturbation  the fixed point  persists as a symmetric
elliptic fixed point of the reversible map $P_{\epsilon}$  --- it is a
point $(0,  y_\epsilon)\in\Gamma_0\cap  \Gamma_1$.  In order to reveal
the nontwist property of the map $P_\epsilon$ we compute and visualize
the  effect  of the map on  different  semi-lines  emanating  from the
elliptic  fixed point  (Figure  \ref{semilines}).  As it results  from
Eq.(\ref{eq:rotation}),  the  rotation  number  for the map  $P_0$  is
negative,   more  exactly   $\rho=-1/\omega$.  Note  that  unlike  the
standard--like  nontwist maps, here a line of constant $\theta$ is not
mapped onto a parabola,  that is a curve with a single  extremum,  but
onto an  oscillating  curve, having many minima and maxima.  This fact
will lead to a different  behaviour of our nontwist map, in comparison
to the dynamical properties of nontwist standard--like mappings.

\begin{figure}[t]
\begin{center}
\includegraphics{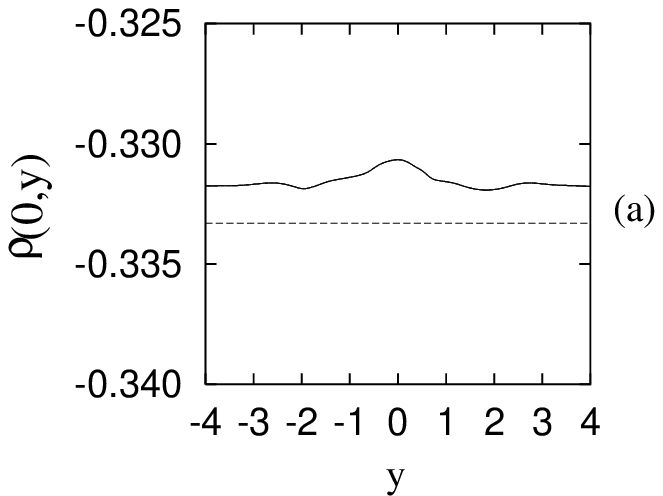}\includegraphics{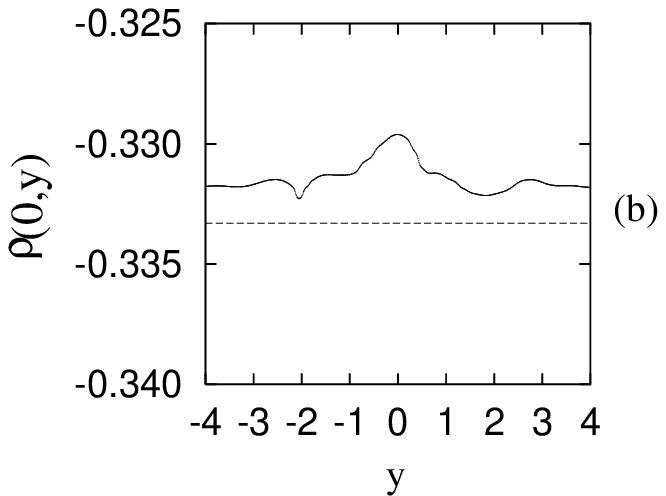}
\includegraphics{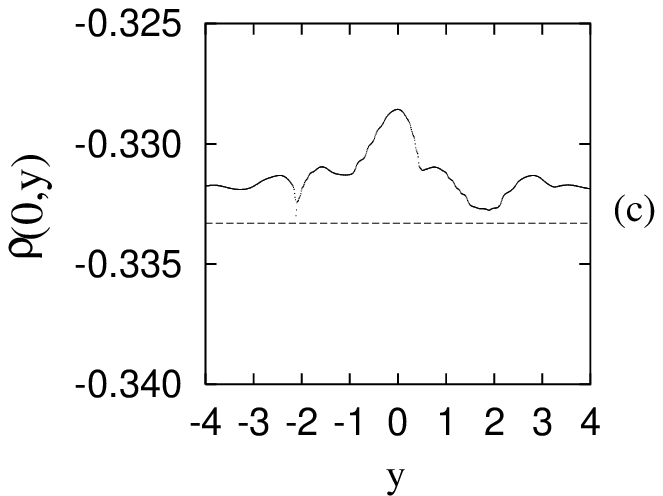}\includegraphics{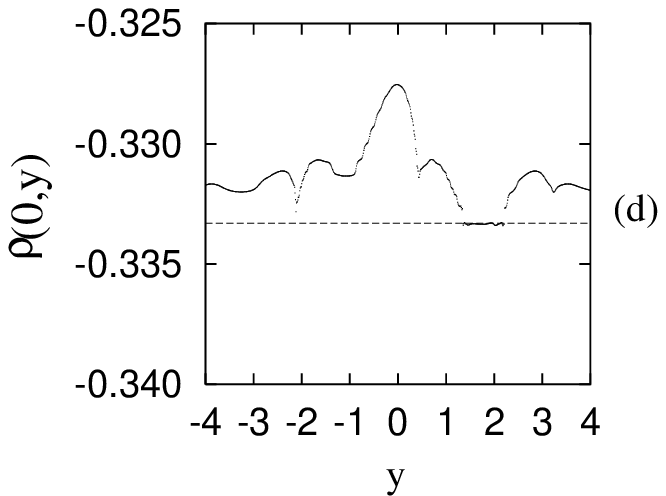}
\includegraphics{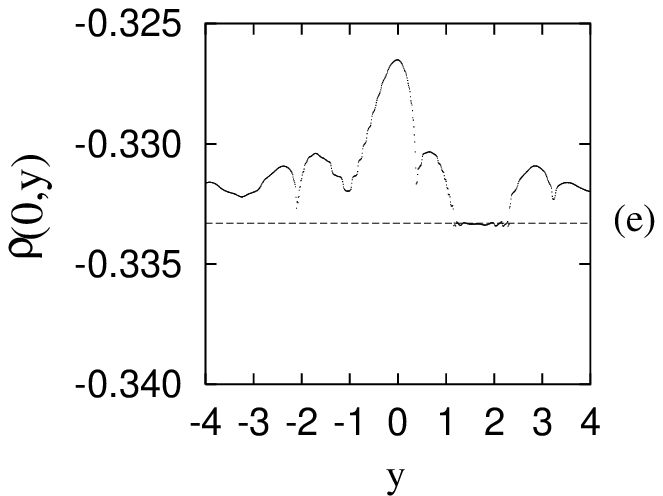}\includegraphics{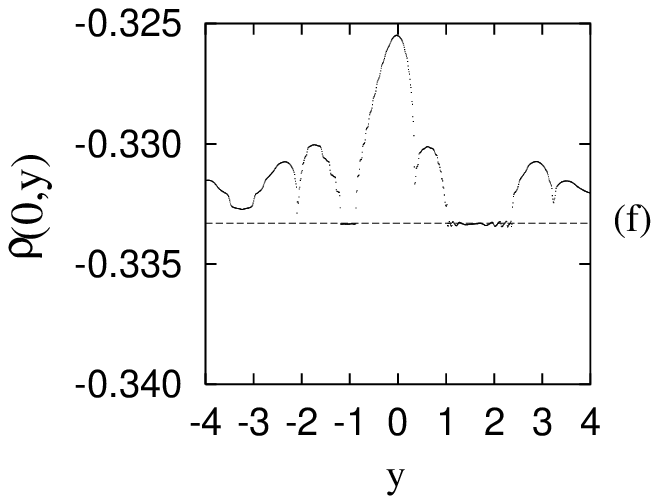}
\end{center}
\caption{The  self-rotation  number of the  orbits of points  $(0,y)$,
	 with  $y\in[-4,4]$.  The  horizontal  axis  is  the  axis  of
	 $y$-values,   while  the  vertical  one  corresponds  to  the
	 self-rotation  number.  The  dashed  line  is  at  the  level
	 $\rho=1/3$.   {\sl   (a)}    $\epsilon=0.01$;    {\sl    (b)}
	 $\epsilon=0.02$;   {\sl  (c)}   $\epsilon=0.03$;   {\sl  (d})
	 $\epsilon=0.04$;   {\sl  (e)}   $\epsilon=0.05$;   {\sl  (f)}
	 $\epsilon=0.06$.}
\label{selfrot}
\end{figure}

\begin{figure}[t]
\centerline{\scalebox{0.9}{\includegraphics{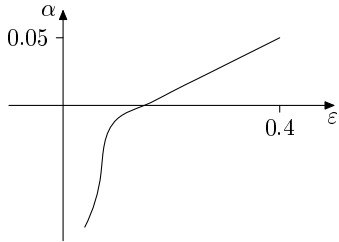}}}
\caption{The   curve   $\alpha^*(\epsilon)$   whose   points  are  the
	 triplication thresholds of the central fixed point of the map
	 $P_\epsilon$.}
\label{bifcurve}
\vspace{0.5cm}
\end{figure}

Besides this  primary way to  illustrate the nontwist property  of the
map $P_\epsilon$, we computed  the self--rotation number of the orbits
of the map $P_\epsilon$, starting  on the symmetry line $\Gamma_0$ and
using  the  turning  angle  method \cite{DEA00}.   The  self--rotation
number for  the map corresponding  to the parameter $\epsilon$  in the
region  of study  is shown  in  Figure \ref{selfrot}.   Hence our  map
$P_\epsilon$  is  a  nontwist  APM,  whose  rotation  number  function
$\rho(y)$ is oscillating, having more than one extremum, that is, more
than one invariant  circle whose rotation number is  a local extremum.
Note that  for the fixed parameter $\omega=3.0146$  of the Hamiltonian
system  of  Eq.(\ref{eq:motion2}),  the  associated  Poincar\'{e}  map
$P_\epsilon$ has an elliptic fixed point which undergoes triplication.
In   Figure  \ref{bifcurve}  we   represent  the   triplication  curve
$\alpha^*(\epsilon)$.

In  order  to  emphasize  the  significant  differences  in the  phase
portraits  before  and after the  triplication  point,  we add that in
Figure \ref{tripl} we illustrated the triplication  bifurcation of the
elliptic  fixed  point  of the  Poincar\'e  map  for  $\omega=3.0146$,
$\epsilon=0.07$   and   $\alpha=-0.3$   ---   panel   {\sl   (a)}  ---
$\alpha=-0.05$  --- panel  {\sl (b)} --- and  $\alpha=0.02$  --- panel
{\sl (c)}.  From the physical point of view it is only of interest the
case in which  $\alpha$  is  constant  and  equal to $\sim  0.3$,  and
$\epsilon$ varies.  Thus, the considered pairs $(\epsilon,\alpha)$ are
located above the triplication  curve.  According to \cite{DMS00}  for
this case a twistless circle can exist.  Comparing the rotation number
associated  to the cubic  H\'{e}non  map  (Figures  \ref{henrot1}  and
\ref{henrot2})  and our Poincar\'{e}  map (Figure  \ref{selfrot}),  we
conclude that the Poincar\'{e} map has an oscillating  rotation number
with a larger  amplitude near the elliptic fixed point.  This could be
interpreted as a  superposition  of the nontwist  property  induced by
triplication   on  the  nontwist   behaviour   due  to  the   periodic
perturbation  of the initial  Hamiltonian  system.  Near the  elliptic
fixed point, the Poincar\'{e}  map has a negative twist.  The apparent
discontinuities  of the rotation number are due to the distance chosen
between the points on $\Gamma_0$  whose  rotation  number is computed.
While on the branch $\Gamma_0^+= \{(x,y)|x=0,y>0\}$, a period--3 orbit
has no point, it has one on the  other  branch,  $\Gamma_0^-=\{(x,y)|x
=0,y<0\}$.  The  orbit  of a  point  on  $\Gamma_0^+$  encircling  the
homoclinic  loops of the hyperbolic  period--3  orbits can lead to an
apparent  discontinuity of the rotation number, although this function
is continuous.  The successive  changes in the monotonicity of the map
lead to a distinct  nontwist  behaviour in  comparison  with the known
results reported for standard--like nontwist maps  \cite{P01,C-NGM96}.

\subsection{Birth  of the dimerized  island  chains.  Local and global
bifurcations}

\begin{figure}[t]
\centerline{\scalebox{0.5}{\includegraphics{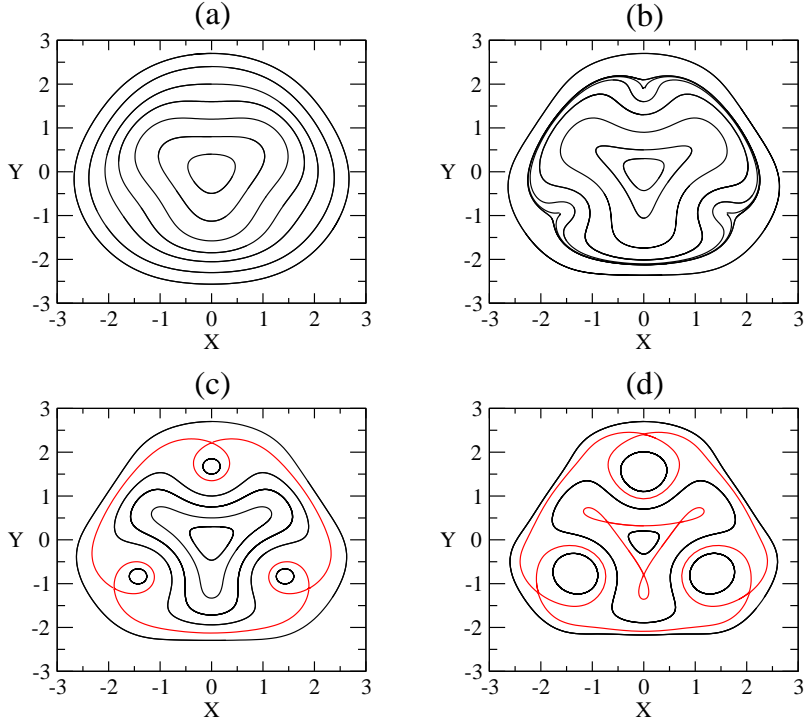}}}  
\caption{The  successive  creation  of the  pair of  dimerized  island
	 chains.  Poincar\'{e}  maps for  $\alpha=0.3$  and  different
	 values  of  $\epsilon$:  {\sl  (a)}  $\epsilon=0.01$  ---  an
	 ``almost'' harmonic oscillator; {\sl (b)} $\epsilon=0.03$ ---
	 the phase portrait just before the saddle-center bifurcation;
	 {\sl  (c)}  $\epsilon=0.04$  --- the  creation  of the  first
	 dimerized  island  chain; {\sl (d)}  $\epsilon=0.07$  --- the
	 pair of period--3  dimerized  island chain is  complete.  The
	 homoclinic  and  heteroclinic  arcs of the island  chains are
	 shown in red.}
\label{dimerized}
\vspace{0.5cm}
\end{figure}

\begin{figure}[t]
\centerline{\scalebox{0.58}{\includegraphics{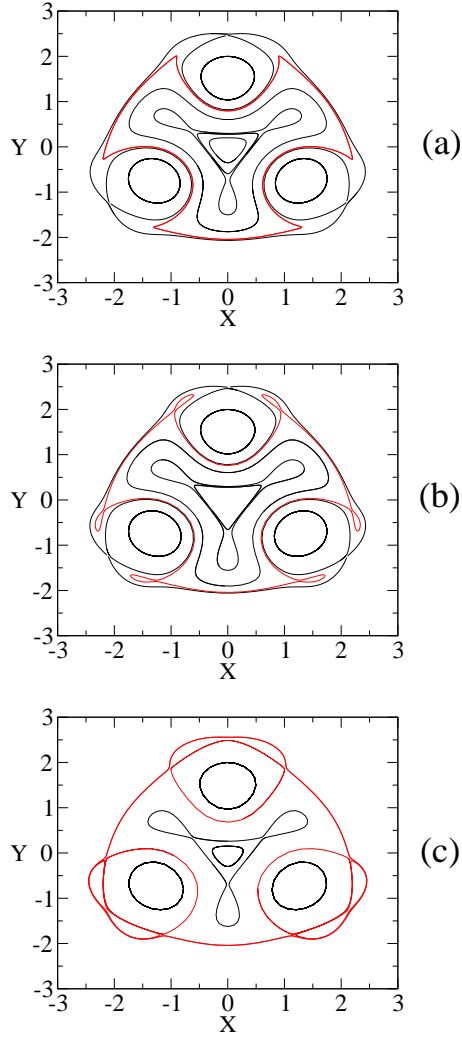}}}
\caption{The    global    bifurcation:   creation   by   saddle-center
	 bifurcations of a pair of period--3  orbits for  $\alpha=0.3$
	 and increasing  values of $\epsilon$.  The particular  orbits
	 involved  in  the  bifurcations  appear  in  red.  {\sl  (a)}
	 $\epsilon=0.0875$  --- the phase portrait at the threshold of
	 a saddle-center  bifurcation.  It presents an invariant curve
	 with six cusps; {\sl (b)} $\epsilon=0.095$ --- two interwined
	 orbits of period--3  have been created.  Inside each new-born
	 loop  there  is  a  period--3   elliptic   point;  {\sl  (c)}
	 $\epsilon=0.11725$   ---   the   threshold   of  the   global
	 bifurcation.  The topology of the separatrices has changed.}
\label{global}
\vspace{0.5cm}
\end{figure}

In our  parametric  study,  we first  considered  the  cases  in which
$\epsilon$ is small, in order to study in detail the  departure of the
system  from the  simple  harmonic  oscillator  as this  parameter  is
increased.  We  restricted  the  study  of the  Poincar\'{e}  map to a
rectangle limited by initial conditions close to physically reasonable
values of the radius and  velocity  of the outer  stellar  layers.  We
focused on the creation of a typical phase-space structure of nontwist
maps:  a  pair  of  dimerized   island  chains.  This   characteristic
scenario of creation of new orbits in nontwist maps is illustrated  in
Figure  \ref{dimerized}.  Panel {\sl a} exhibits the phase portrait of
the  near--to--integrable  map,  where  the  elliptic  fixed  point is
surrounded  by  invariant  circles.  At  $\epsilon  \approx  0.03$ the
Poincar\'{e}  map has an  invariant  curve with  three  cusps  (Figure
\ref{dimerized}{\sl b}).  Each cusp is a point of a new periodic orbit
created  through a  saddle-center  bifurcation  for $\epsilon  \approx
0.04$:  a stable and an unstable  fixed point  emanate  from each cusp
(Figure   \ref{dimerized}{\sl   c}).  The  homoclinic  loop  of  every
hyperbolic  point surrounds the  corresponding  elliptic one.  Between
two  successive  hyperbolic  points  (in  cyclic  order)  there  is  a
heteroclinic  connection.  This separatrix  structure called dimerized
island  chain  appears  in red in Figure  \ref{dimerized}{\sl  c}.  In
Figure  \ref{selfrot}{\sl  d} we have shown the  self-rotation  number
$\rho(0,y)$  for the case  whose  Poincar\'e  map  appears  in  Figure
\ref{dimerized}{\sl  c}.  In the former figure, the plateau $\rho(0,y)
=-1/3$ corresponding to $y \in (1,2)$ represents the constant rotation
number due to the newly created elliptic fixed point on  $\Gamma_0^+$.
The  dimerized  island  chain is also  responsable  for the minimum of
$\rho(0,y)$ around $y \approx -2$.

Generically, dimerized island chains of the same period exist in pairs
and are born  in stages.  The second period--3  dimerized island chain
is created through  the same bifurcation process at  a higher value of
$\epsilon$    (Figure     \ref{dimerized}{\sl    d}).     In    Figure
\ref{selfrot}{\sl  f},  the small  plateau  of  $\rho(0,y)$ around  $y
\approx -1$ explains  the birth of the second  dimerized island chain.
Between the two island chains the invariant circles are meanders, that
is, the  radius along such a circle  is not a univoque  function of the
angle.

Around the last  born period--3 dimerized island chain,  a sequence of
local and  global bifurcations occurs.  This is  illustrated in Figure
\ref{global}, where it can be  seen that two independent orbits of the
same period--3 are created  by saddle-center bifurcation.  They evolve
in  such a  way that,  finally, they  interact with  the  orbits which
belong to  the first dimerized island chain.   As $\epsilon$ increases
the newly born  elliptic points approach the hyperbolic  points of the
dimerized chain.  When $\epsilon$ reaches  a value of 0.11725 a global
bifurcation  occurs:  the  newly  created orbits  interfere,  and  the
hyperbolic  points of  the  dimerized island  chain become  hyperbolic
points with homoclinic eight-like  orbits encircling the newly created
elliptic points.

\begin{figure}[t]
\centerline{\scalebox{0.43}{\includegraphics{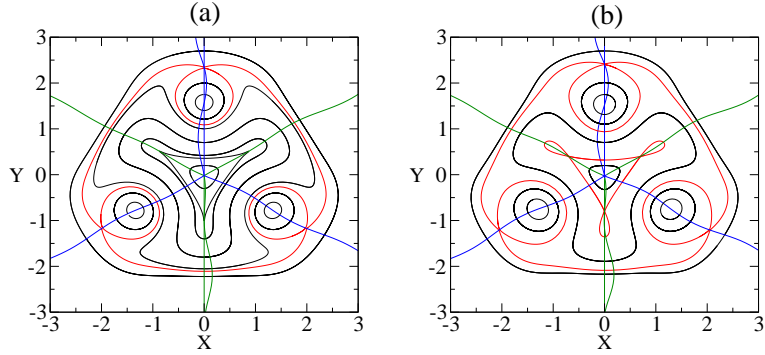}}}
\caption{The   Poincar\'{e}  map  $P_\epsilon$   before  and  after  a
	 saddle-center  bifurcation.  The  figures  display  also  the
	 symmetry lines  $\Gamma_{2k}^+$ (in blue) and $\Gamma_{2k}^-$
	 (in green), with $k$=0,1,2,3.  {\sl (a)} $\epsilon=0.055$ ---
	 the  phase  portrait  prior  to the  creation  of the  second
	 dimerized  island  chain; {\sl (b)}  $\epsilon=0.07$  --- the
	 second  dimerized  island chain has been created.  The points
	 satisfying $(x,y) \in \Gamma_0^{\pm} \cap \Gamma_6^{\pm}$ are
	 period--3 points.}
\label{symlines}
\vspace{0.5cm}
\end{figure}

In  order  to show  the  successive  births  of  period--3  orbits  as
$\epsilon$ increases with $\alpha=0.3$ fixed, we compute and visualize
$P_\epsilon^k (\Gamma_0^+)$ and $P_\epsilon^k(\Gamma_0^-)$, $k=1,2,3$.
This process was first  illustrated in Figure  \ref{semilines}{\sl  a}
where  $\Gamma_0^+$ and $P_\epsilon  ^1(\Gamma_0^+)=  \Gamma_2^+$ were
represented.  We show in Figure  \ref{symlines} the connection between
the  symmetry  properties  of the map and  the  creation  of  periodic
orbits.  A  saddle-center  bifurcation  occurs  at  the  value  of the
parameter $\epsilon$ at which  $\Gamma_6^\pm$ has a tangential contact
with $\Gamma_0^\pm$.  The corresponding  Poincar\'{e} map has for such
an $\epsilon$ an invariant  curve with cusps which  represent  the
points of tangency.  Increasing  the  perturbation,  the two  symmetry
lines  intersect  at two  points,  one being  elliptic  and the other,
hyperbolic.

\subsection{Creation of chains of vortices}

\begin{figure}[t]
\centerline{\scalebox{0.57}{\includegraphics{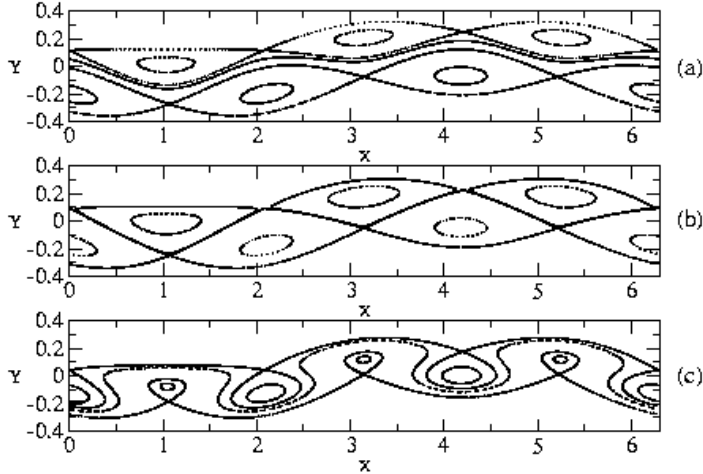}}}
\caption{Reconnection    process   in   the   nontwist   standard-like
	 multi-harmonic  map.  {\sl  (a)}   $\omega=0.3287$   ---  two
	 independent    Poincar\'e-Birkhoff   chains   ;   {\sl   (b)}
	 $\omega=0.3297655$ --- the reconnection  threshold; {\sl (c)}
	 $\omega=0.3315$   ---  two   dimerized   island   chains.  We
	 illustrate also a meander separating the two dimerized island
	 chains.}
\label{multiham}
\vspace{0.5cm}
\end{figure}

A  global  bifurcation  which  is  generic  of  nontwist  maps  is the
so-called   reconnection  process,  which  is  illustrated  in  Figure
\ref{multiham}   for  the   nontwist   multi-harmonic   standard   map
\cite{P01}:

\begin{eqnarray}
x_{n+1}&=&     x_n+2\pi\omega+y_{n+1}^2~\,\,(\mbox{mod}\,\,    2\pi)\\
y_{n+1} & = & y_n+ k\sin [x_n + \arcsin (e \sin x_n)],
\end{eqnarray}

\noindent with $k=0.2$, $e=0.38$ and different values of $\omega$.  On
the two sides of a twistless circle, two Poincar\'{e}--Birkhoff chains
(necklaces of consecutive  elliptic and hyperbolic points), having the
same rotation number,  approach each other (Figure  \ref{multiham}{\sl
a}).  At  a  given  threshold,   which  is  called  the   reconnection
threshold, their hyperbolic points are connected by heteroclinic  arcs
(Figure  \ref{multiham}{\sl  b}).  Varying the parameter of the system
further, two dimerized  island  chains  emerge from the  configuration
created by the  reconnection  (Figure  \ref{multiham}{\sl  c}).  These
chains are separated by meanders.

\begin{figure}[t]
\centerline{\scalebox{0.43}{\includegraphics{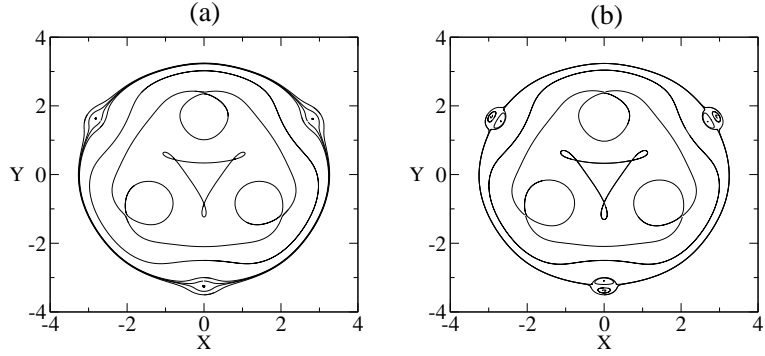}}}
\caption{Creation  of the first chain of  vortices  for  $\alpha=0.3$.
	 {\sl (a)} $\epsilon  =0.0614$ The  saddle-center  bifurcation
	 has  created  the first  dimerized  island  chain;  {\sl (b)}
	 $\epsilon=0.06617$  --- instead of a second  dimerized island
	 chain,  a pair  of  vortices  is  born.  See  text  for  more
	 details.}
\label{vortices1}
\vspace{0.5cm}
\end{figure}

In our  case, we  witness the creation  of the nongeneric  vortices or
dipoles \cite{C-NGM96}.  Figure  \ref{vortices1} displays the birth of
the  first  chain  of   vortices.   As  $\epsilon$  increases,  either
$\Gamma_6^+$ or  $\Gamma_6^-$ intersects  tangentially at a  new point
located  at  a  radius  larger  than  in the  previous  cases.   As  a
consequence, new  period--3 dimerized island chains  are created after
$\epsilon$  crosses the  value of  tangential contact.   However, each
second  dimerized island  chain  has a  different creation  mechanism.
Unlike the previous case, the  hyperbolic point of the first dimerized
island chain bifurcates into  two hyperbolic points in the transversal
direction.   In this  process, a  new elliptic  point is  born  on the
symmetry  line.   These  hyperbolic  points  are  connected  by  three
heteroclinic  arcs: one surrounding  the previously  existing elliptic
point, another surrounding  the new elliptic point, and  the third one
separating  the two  elliptic points.   Thus,  a pair  of vortices  is
created.

\begin{figure}[b]
\vspace{0.8cm}
\centerline{\scalebox{0.35}{\includegraphics{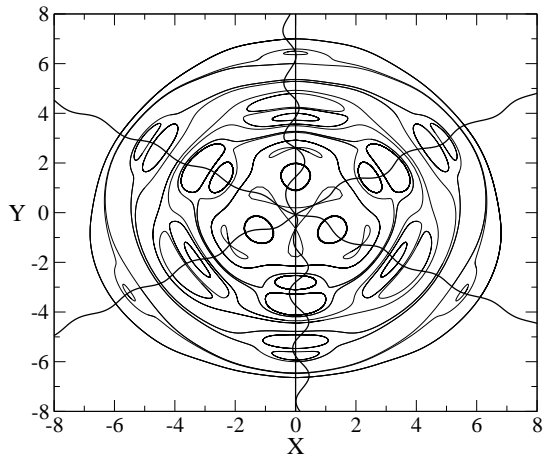}}}
\caption{The  phase portrait of the Poincar\'{e} map  corresponding to
	 $\epsilon=0.2$  and  $\alpha=0.3$.  One can see a sequence of
	 concentric  chains  of  vortices.  The most  exterior  one is
	 still  incomplete.  The  figure  displays  also the  symmetry
	 lines $\Gamma_{2k}^{\pm}$, $k$=0,1,2,3.}
\label{vortices2}
\vspace{0.5cm}
\end{figure}

As  $\epsilon$  increases,  the  process  of formation  of  chains  of
vortices  continues for larger  radii (Figure  \ref{vortices2}).  Note
that  if  the elliptic  orbits  of  one  chain of  vortices  intersect
$\Gamma_0^+$, then the next one,  which is created external to it, has
a  pair of  elliptic points  on the  symmetry line  $\Gamma_0^-$.  The
invariant rotational circle interpolating the hyperbolic points of the
chain of vortices is  the twistless circle.  In Figure \ref{vortices2}
it can be  seen that such circles pass almost  through the extremum of
the symmetry lines $\Gamma_6^\pm$.   As already said, the formation of
vortices  is  not generic  and  therefore  the  twistless circle  that
appears between them is not either.

\subsection{Formation of the stochastic sea}

\begin{figure}[t]
\centerline{\scalebox{0.58}{\includegraphics{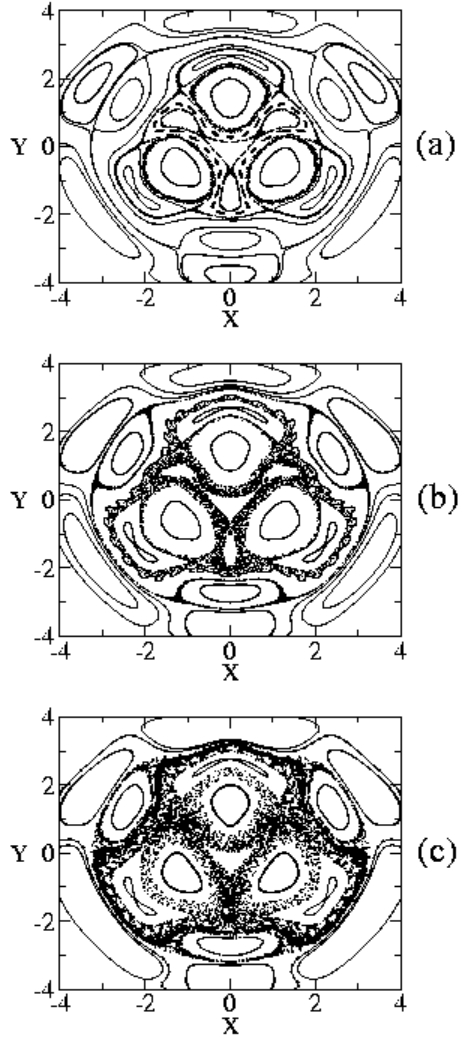}}}
\caption{Formation  of the  stochastic  sea($\alpha=0.3$):  {\sl  (a)}
	 $\epsilon=0.3$  --- as the second  dimerized  island chain is
	 destroyed,  it is clothed in a  stochastic  layer;  {\sl (b)}
	 $\epsilon=0.4$  --- the stronger the perturbation,  the wider
	 the   stochastic   layers;  {\sl  (c)} $\epsilon=0.5$  ---  
         merging  of  the
	 stochastic layers forming the stochastic sea.}
\label{chaotic}
\vspace{0.5cm}
\end{figure}

Our system  is a clear example  of weak chaos,  where the perturbation
itself  creates a separatrix  network at  a certain  $\epsilon_0$ (the
dimerized island chains and the  vortices chains) and then destroys it
as $\epsilon$ increases by producing regions of chaotic dynamics.  The
case  of  strong  chaos   implies  that  the  unperturbed  Hamiltonian
intrinsically  has  separatrix structures  and  the perturbation  just
clothes them in thin stochastic layers.  In both cases, the merging of
the stochastic  layers in the  phase space may  give rise to  a single
chaotic network called stochastic  web.  The larger $\epsilon$ is, the
wider  is the  stochastic web.   Inside the  cells of  the  web, there
exists a  set of islands  of regular motion called  invariant web-tori
\cite{ZEA91}  whose  dimensions  are  inversely  proportional  to  the
perturbation strength: for strong  perturbations, they are engulfed by
the stochastic sea. Our system  presents this kind of behaviour, which
is presented in Figure \ref{chaotic}.

\begin{figure}[t]
\centerline{\scalebox{0.43}{\includegraphics{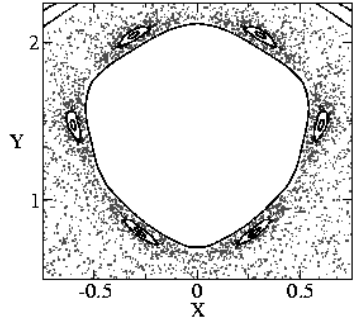}}}
\caption{A  zoom  of  Figure  \ref{chaotic}{\sl  b}  illustrating  the
         islands-around-islands hierarchy.}
\label{particular}
\end{figure}

It  has  long  been known  (and  not  yet  well understood)  that  the
boundaries of the islands of regular motion can be ``sticky'', meaning
that chaotic  orbits can spend  a long time  close to the  islands and
then escape in the chaotic  sea.  This phenomenon was observed already
in \cite{C71}  and was later called {\sl  stickiness} \cite{SR82}.  In
general, it  can be  explained by the  transformation of  the web-tori
into web-cantori  as the perturbation  is increased.  Contrary  to the
absolute barriers  constituted by the invariant tori,  the cantori are
only  temporary barriers.   That  is,  the orbits  close  to the  main
islands can reach the chaotic sea  when they encounter the gaps in the
cantori,  which happens  after  an indefinite  time interval.   Recent
works \cite{RZ99,ZEN97} also suggest that the stickiness is due to the
formation of  higher-order resonant  islands in the  transition region
between the main island and the chaotic sea (Figure \ref{particular}).
The presence  of these so-called  {\sl dynamical traps} appears  to be
generic  for   Hamiltonian  systems  and   therefore  their  existence
critically determines  the large-timescale behaviour  of such systems.
In this  context, a behaviour  extremely similar to sticky  orbits has
been  shown to result  from simulations  of stellar  variability using
full    hydrodynamical   codes    carried   on    significantly   long
time-intervals.   In  the absence  of  a  better  explanation, it  was
attributed  to long-term  (secular) nonlinear  effects in  the stellar
envelopes.  In  a previous work  \cite{MEA02}, we have shown  that our
simple model recovers such a behaviour as it is generic of Hamiltonian
systems.  We  also presented clear examples of  sticky orbits together
with an extensive discussion of their implications in the framework of
the classification of variable stars.  Therefore, we mention here only
the fact that if a Hamiltonian approach is intrinsic to the phenomenon
of  stellar variability, we  can expect  that different  categories of
variable  stars (classified according  to their  amplitude, frequency,
irregularity) may be viewed under  a more unifying perspective than it
is  presently  considered.  Moreover,  considering  that the  sporadic
excursions of the sticky orbits to the chaotic sea translate into high
velocities of  the outer  layers, mass loss  is very likely  to occur,
again in accordance with the observations.

\begin{figure}[t]
\centerline{\scalebox{0.43}{\includegraphics{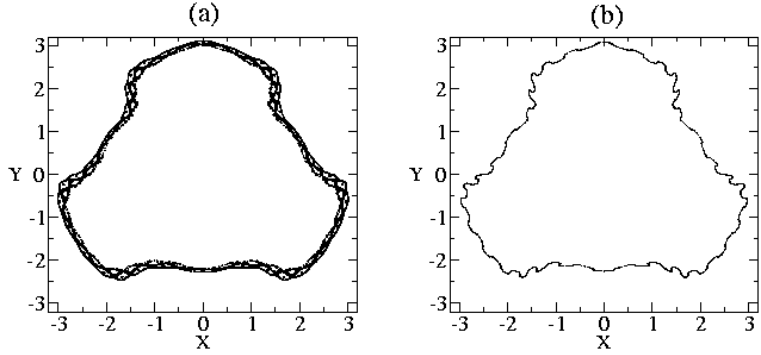}}}
\caption{The  reconnection process:  {\sl (a)} $\epsilon=0.405$ -- The
	 Poincar\'e-Birkhoff  chains confirming the generic process of
	 reconnection  illustrated in Figure \ref{multiham}; {\sl (b)}
	 $\epsilon=0.4$  --  After  the  reconnection,  the  resultant
	 meander   with  34  cusps,   slightly   visible   in   Figure
	 \ref{chaotic}{\sl b}.}
\label{meander}
\vspace{0.5cm}
\end{figure}

Another  important and  distinctive  feature of the phase  portrait of
nontwist maps is the existence of {\sl  meanders}, as it was mentioned
before when discussing  Figure  \ref{dimerized}{\sl  d}.  Meanders are
created  between two  successively  born  dimerized  island  chains or
between  two chains of  vortices.  In nontwist  standard--like  maps they
become  usual  invariant  curves  after  the  reconnection  of the two
chains.  For the sake of clarity, we illustrate next the  relationship
between the  meanders  and the  reconnection  process for our map.  In
Figure  \ref{meander}  we show  the  creation  of  meanders  from  the
reconnection of the  Poincar\'e-Birkhoff  chains of period 34.  In the
case of the standard nontwist map  \cite{C-NGM96}, in the reconnection
of Poincar\'e-Birkhoff chains of even periodic orbits, periodic points of
the chains  having the same  stability  type are aligned in phase.  In
other  words,  to  an  elliptic   (hyperbolic)   point  of  one  chain
corresponds below or above also an elliptic  (hyperbolic)  point.  Due
to this alignment, the reconnection is nongeneric and corresponds to a
hyperbolic-hyperbolic  collision leading to the formation of vortices.
In our case, the two  Poincare-Birkhoff  chains of even period (period
34) have periodic points of opposite  stability type which  approach each
other in their  way to  reconnection  (Figure  \ref{meander}{\sl  a}).
Therefore, the reconnection is generic and in the subsequent dimerized
islands chains, hyperbolic-elliptic collision occurs.  Moreover, after
the  reconnection,  the  twistless  circle  turns  from a  graph  of a
function   of   the   angular    variable   to   a   meander   (Figure
\ref{meander}{\sl  b}).  This  meander is  slightly  visible in Figure
\ref{chaotic}{\sl b}.

Meanders  appear  to  be  robust  invariant  circles  acting  also  as
separatrices  between  distinct  regions  of  chaotic  dynamics.   For
stronger perturbations, their destruction allows the chaotic orbits to
reach other  stochastic regions  previously separated by  the meander.
This behaviour was observed in nontwist standard-like maps \cite{S98},
but until now there is no explanation for this robustness.

\section{Conclusions}

In  the present  work  we  have extended  the  preliminary results  of
Ref. \cite{MEA02} concerning the dynamics  of a forced oscillator as a
model of  irregular stellar pulsations.  The  driving is characterized
by  two   parameters,  the   fractional  amplitude  of   the  internal
perturbation  $\alpha$   and  the  total  amplitude   of  the  driving
$\epsilon$.  Our aim was to  undertake a more detailed analysis of the
transition to the nontwist  property in the associated Poincar\'e map.
We have  demonstrated that for a  given $\epsilon >0$  there exists an
$\alpha^*(\epsilon)$ that corresponds to a triplication of the central
elliptic fixed point  of the map.  As we were  interested in values of
the parameters  $\alpha$ and $\epsilon$ above  the triplication curve,
the nontwist character of the map was undoubtly present \cite{DMS00}.

Special  attention was devoted to the process of formation of periodic
orbits and to  periodic-orbits  collision.  The  reversibility  of the
Poincar\'e map and  consequently,  the symmetry  properties  allowed a
clear  identification  of the structure of the periodic  points.  More
precisely, the evolution and the  intersections  of the symmetry lines
of the  system  provided  good  tracers  of the steps of  creation  of
dimerized  island  chains and vortex  chains.  We  strongly  argued in
favor of these  particularities  of the map being  entirely due to the
superposition  of two factors:  the  nonmonotonicity  of the  rotation
number  function  induced by the  triplication  of the elliptic  fixed
point  and  the  nonmonotonicity  of  the  same  function  due  to the
oscillating   perturbation.  The  latter  feature  admits  alternative
occurrence  of  maxima  and  minima,  where  the  twist  condition  is
violated.  This translates  into the minima and maxima of the symmetry
lines which transform the invariant  curves passing  through them into
twistless  circles.  Using these  elements, we have followed the local
and global  bifurcations  until the  formation  of  stochastic  layers
around  separatrices  together with the associated  sticky orbits.  We
have proved that the  reconnection of even periodic  orbits is generic
by following the  reconnection  of two  Poincar\'e-Birkhoff  chains of
period 34.  From the  astrophysical  point of view, we have dwelled on
the  implications  of the sticky regions and of the interplay  between
stochastic  and regular  regions in the  framework of the  analysis of
stellar    variability.   In    order    to    provide    quantitative
characterization   of  sticky  orbits  in  the  framework  of  stellar
variability,  we shall  concentrate  in the  future on the  causes and
probabilistic estimates of the stickiness mechanism.

\vspace{0.2cm}

\noindent {\sl Acknowledgements.}  This work has been supported by the
DGES grant  PB98--1183--C03--02, by the MCYT  grant AYA2000--1785, and
by the CIRIT  grant 1999SGR-00257.  We also would  like to acknowledge
many helpful discussions with G.M.  Zaslavsky.


\begin{thebibliography}{00}

\bibitem{M92}      J.D.  Meiss,   Rev.  Mod.  Phys.,   {\bf  64},  795
		   (1992)
\bibitem{C-NEA97}  D.  del   Castillo-Negrete,   J.M.   Greene,   P.J.
		   Morrison, Physica D, {\bf 100}, 311 (1997)
\bibitem{SA97}     S.  Shinohara,   Y.  Aikawa,   Prog.  Theo.  Phys.,
		   {\bf 97}, 379 (1997)
\bibitem{P01}      E.  Petrisor,  Int.  J.  Bif.  \& Chaos,  {\bf 11},
		   497 (2001)
\bibitem{IEA92}    V.  Icke, A.  Frank, A.  Heske,  A\&A,  {\bf  258},
		   341 (1992)
\bibitem{MEA02}    A.  Munteanu,  E.  Garc\'{i}a-Berro,  J.  Jos\'{e},
		   E.  Petrisor, Chaos, {\bf 12}, 332 (2002)
\bibitem{REA96}    C.  Ritossa,   E.   Garc\'\i   a-Berro,   I.  Iben,
		   Astrophys.  J., {\bf 460}, 489 (1996)
\bibitem{GEA97}    E.  Garc\'\i   a-Berro,   C.   Ritossa,   I.  Iben,
		   Astrophys.  J., {\bf 485}, 765 (1997)
\bibitem{C-NGM96}  D.    del  Castillo-Negrete,  J.M.    Greene,  P.J.
                   Morrison, Physica D, {\bf 91}, 1 (1996)
\bibitem{S98}      C.  Sim\'{o}, Regular \& Chaotic Dyn., {\bf 3}, 180
		   (1998)
\bibitem{C-NF02}   D.  del Castillo-Negrete,  M.C.  Firpo, Chaos, {\bf
		   12}, 496 (2002)
\bibitem{HH95}     J.E.  Howard, J.  Humpherys,  Physica  D, {\bf 80},
		   256 (1995)
\bibitem{SEA97}    S.   Sait\^{o},    Y.   Nomura,   K.   Hirose,   H.
		   Ichikawa, Chaos, {\bf 7}, 245 (1997)
\bibitem{DMS00}    H.R.    Dullin,    J.D.   Meiss,    D.    Sterling,
		   Nonlinearity, {\bf 13}, 203 (2000)
\bibitem{L98}      J.S.W. Lamb, Physica D, {\bf 112}, 1 (1998)
\bibitem{DEA00}    H.R.  Dullin, J.D.  Meiss, D.  Sterling, Physica D,
		   {\bf 145}, 25 (2000)
\bibitem{ZEA91}    G.  Zaslavsky,  R.Z.  Sagdeev,  D.A.  Usikov,  A.A.
		   Chernikov,   {\sl  Weak  chaos  and   quasi-regular
		   patterns},   Cambridge   Univ.   Press:   Cambridge
		   (1991)
\bibitem{C71}      G. Contopoulos, Astron. J., {\bf 76}, 147 (1971)
\bibitem{SR82}     R.B.  Shirts,  W.P.  Reinhardt,   J.  Chem.  Phys.,
		   {\bf 77}, 5204 (1982)
\bibitem{RZ99}     V.  Rom-Kedar, G.M.  Zaslavsky, Chaos, {\bf 9}, 697
		   (1999)
\bibitem{ZEN97}    G.  Zaslavsky, M.  Edelman,  B.A.  Niyazov,  Chaos,
		   {\bf 7}, 159 (1997)

\end{thebibliography}
\end{document}